\documentclass[aps,pre,twocolumn,groupedaddress,amsmath,amssymb]{revtex4-1}
\usepackage{graphicx}  % needed for figures
\usepackage{dcolumn}   % needed for some tables
\usepackage{bm}        % for math
\usepackage{verbatim}   % for math
\usepackage{epstopdf}
\usepackage{subcaption}					% for sub-figures
\usepackage{color}
\definecolor{orange}{rgb}{1,0.5,0}
\definecolor{brown}{rgb}{0.65, 0.16, 0.16}
\definecolor{phlox}{rgb}{0.87, 0.0, 1.0}
\graphicspath{{figs/}}					% the path of image folder

\begin{document}
	
	\title{Analytical approach to the surface plasmon resonance characteristic of metal nanoparticle dimer in dipole-dipole approximation}
	\author {Asef Kheirandish}
	\affiliation{Department of Physics, University of Mohaghegh Ardabili, P.O. Box 179, Ardabil, Iran}
	\author {Nasser Sepehri Javan}
	\email{nsj108119@yahoo.com}
	\affiliation{Department of Physics, University of Mohaghegh Ardabili, P.O. Box 179, Ardabil, Iran}
	\author {Hosein Mohammadzadeh}
	\affiliation{Department of Physics, University of Mohaghegh Ardabili, P.O. Box 179, Ardabil, Iran}

	\pacs{51.30.+i,05.70.-a}
	
		\begin{abstract}
		This theoretical study, deals with the effect of bi-particle interaction on the surface plasmon resonance (SPR) in a dimer which includes two identical metal nanoparticles (NPs). Considering the dipole-dipole interaction in a Drude-like model, an appropriate equation is derived for the permittivity of each NP. The restoration force related to the classical confinement originating from the finite size of NPs is considered and an appropriate adjustment coefficient is considered for this term through analyzing experimental data. Two different polarizations are considered for the laser beam electric field and it is shown that the orientation of electric field has essential role in the linear optical properties of dimer. Numerical investigation is accomplished for a dimer of gold NPs with two different diameters of 4nm and 20nm. For the parallel polarization, dipole-dipole interaction leads to the red-shift of SPR wavelength and increase in its peak value while for the perpendicular polarization, the absolutely opposite results are derived. For all cases, it is shown that SPR wavelength functionality with respect to the geometric factor $a/d$ (NP radius to the separation) can be presented by a cubic equation that fits better than an exponential one suggested by the earlier studies which demonstrates the dipole-dipole characteristic of the interaction. Qualitatively, our results are in good agreement with the other experimental studies.\\\\
		{\bf Keywords:} laser, interactional nanoparticles, plasmon, redshift, blueshift, dipole-dipole interaction.
	\end{abstract}
	
	\maketitle

	%%%%%%%%%%%%%%%%%%%%%%%%%%%%%%%%%%%%%%%%%%%%%%%%%%%
	\section{Introduction}
	%%%%%%%%%%%%%%%%%%%%%%%%%%%%%%%%%%%%%%%%%%%%%%%%%%%
	There is a progressive interest in the theoretical and experimental studies of interaction of electromagnetic waves with metallic nanostructures. In addition to the attractive physics, cutting-edge technological applications of light-NPs interaction caused a doubled motivation for scientists to intensively study the linear and non-linear properties of metal nanoparticles (MNPs). The main aspect in the light-MNPs interaction is the SPR which refers to the resonant excitation of collective oscillations of conduction free electrons of particle in the interaction with the light electric field \cite{garcia2011surface} where the linear and non-linear properties of MNPs substantially are improved. The SPR leads to a huge enhancement of electric near-field in the vicinity of the MNP surface, which forms the foundation of surface-enhanced spectroscopy \cite{chen1983surface,schatz1984theoretical}. Besides the vital applications of SPR behavior in the biomedicine \cite{fritzsche2003metal,hu2006gold,eustis2006gold,garcia2011surface}, energy \cite{pillai2007surface,kelzenberg2010enhanced,matheu2008metal}, environment science \cite{narayanan2005catalysis,awazu2008plasmonic,larsson2009nanoplasmonic}, sensing \cite{homola2006surface} and information technology \cite{ozbay2006plasmonics,maier2006plasmonics,barnes2003surface} this effect provides a unique opportunity to determine the kind, shape and size of MNPs based on the resonance frequency \cite{ditlbacher2000spectrally,haiss2007determination,myroshnychenko2008modelling}.\\
	Presence of another NP in the neighborhood can substantially change the SPR via induced multipolar fields. In some recent experimental investigations, dominant dipole-dipole interparticle interaction evidences are reported for nanostructured dimers of MNPs \cite{halas2011plasmons,sheikholeslami2010coupling,haynes2003nanoparticle}. For very close placement of particles in a dimer, relevantly the dipole-dipole nature of interaction cannot be realistic and multi-polar coupling together with quantum tunneling effect should be considered \cite{ghosh2007interparticle,romero2006plasmons,scholl2013observation}. The effects of interparticle separation, shape and kind of MNP and surrounding medium on the plasmon resonance of dimers are experimentally studied \cite{halas2011plasmons,sheikholeslami2010coupling,kelly2003optical,hooshmand2015plasmonic}. For large separation of MNPs dimers, the exponential behavior of SPR peak shift has been proved in some experimental studies of El-Seyed and coworkers \cite{su2003interparticle,tabor2008use,jain2007universal}. In a very recent simulation and experimental study for cubic and spherical silver dimers \cite{hooshmand2019collective}, it is found that for the cubic NPs dimer, even in the small sizes of NPs, the contribution of the higher-order modes is remarkable because of the appearance of the high density of electric dipoles intensified on the cube corners.\\
	In this theoretical study, we derive analytical expressions for the permittivity of MNPs dimer including two identical spherical particles. Permittivity of each MNP is obtained via a Drude-like model were the restoration force caused by the limited size of particle is included and for the interaction of particles, the dipole-dipole approximation is considered. Two different orientations of laser beam electric field with respect to the dimer axis are studied. Numerical experiments are accomplished for different sizes of gold NP dimers. Increase in the resonance peak of extinction cross section and its red-shift are observed for the parallel polarization while absolutely opposite results are obtained for the perpendicular orientation of laser beam electric field which confirm the existing experimental results. Even though the exponential function can express the functionality of SPR wavelength with respect to the ratio of radius to the separation as it has been shown in literature \cite{jain2007universal,hooshmand2019collective}, however dipole-dipole property of interaction can be stated by a cubic equation more accurately.

%********************************************
\section{PLASMON COUPLING IN METAL NANOPARTICLE DIMER}
%********************************************

%********************************************
\subsection{Parallel Polarization}
%********************************************

Let us consider interaction of an incident linearly-polarized laser beam with a spherical NPs system which contains two identical MNPs with radius $a$ separated by the distance $d$   according to Fig. (1-a). The laser beam electric filed is taken parallel to the orientation of dimer axis (i. e., orientation of the line segment bounded between two centers of NPs) as following
\begin{align}
{{\bf{E}}_L} = \frac{1}{2}E{e^{i(kx - \omega t)}}{{\bf{\hat e}}_z} + c.c.,
\end{align}
where $E$, $k$ and $\omega$ are the amplitude, wavenumber and frequency of the incident wave and $c.c.$ denotes the complex conjugate.\\
For the NP whose radius is very smaller than the wavelength, we can neglect the variations of the electric field inside the NP and suppose that all electrons experience the same force at a given time where the used approach is called the rigid body approximation. In this case, all electrons behave identically and move the same distance from the rest state. Because of the symmetry of problem, dynamics of both NPs is the same, therefore, first, we consider only the dynamics of one of them.\\
\begin{figure}[t]
	\begin{subfigure}{0.35\textwidth}\includegraphics[width=\textwidth]{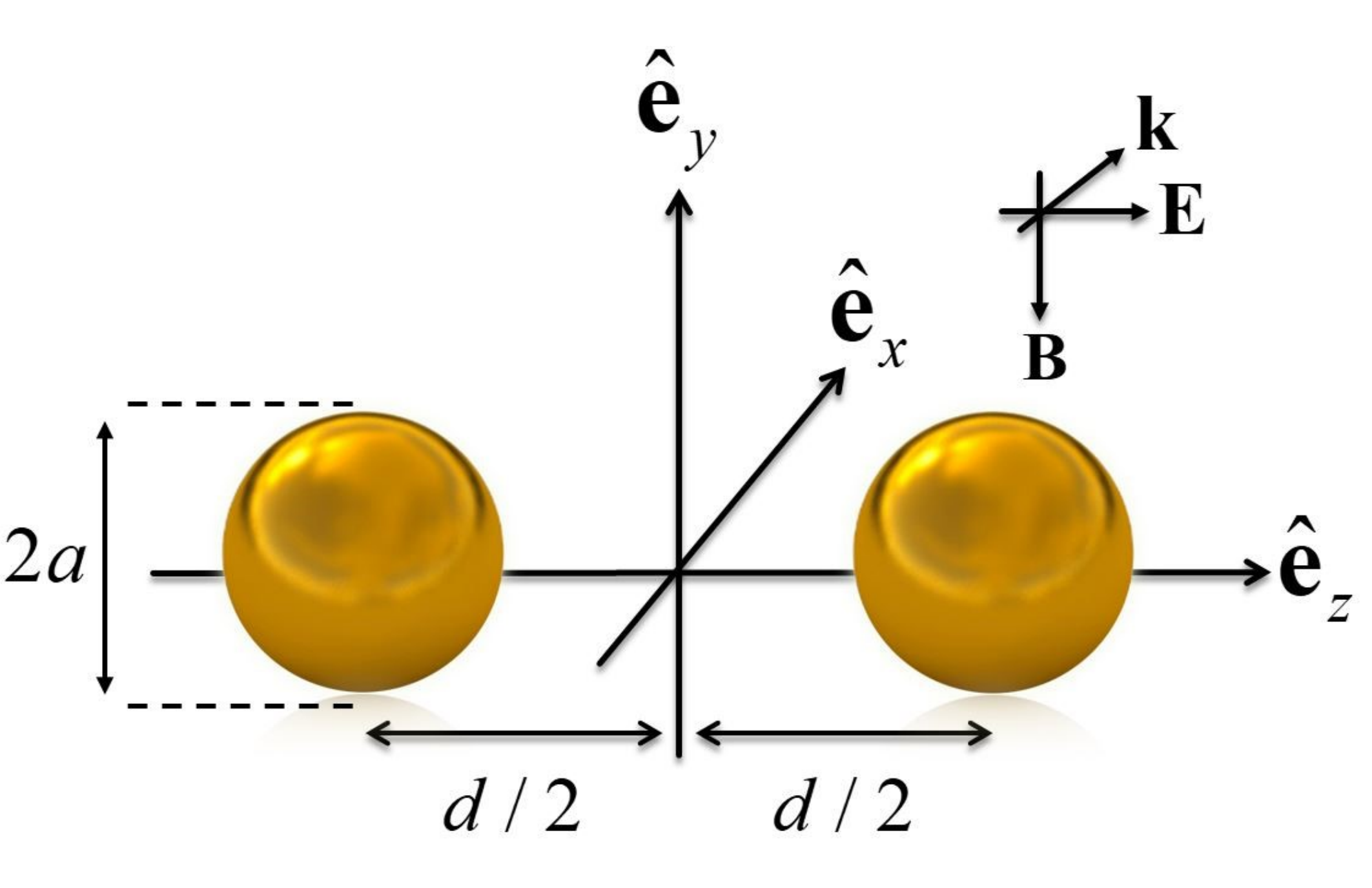}
		\caption{}
		\label{fig1a}
	\end{subfigure}
	\begin{subfigure}{0.35\textwidth}\includegraphics[width=\textwidth]{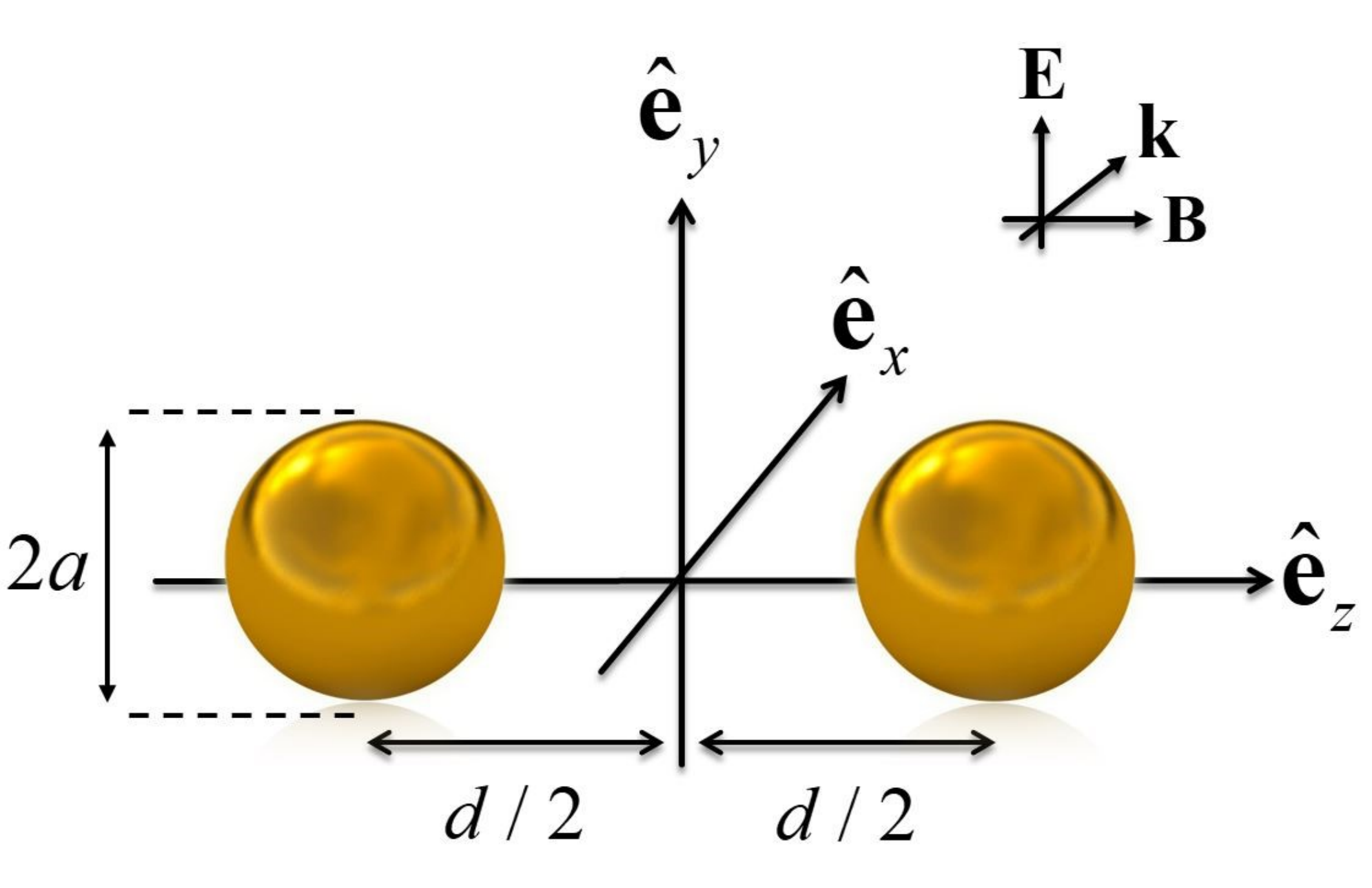}
		\caption{}
		\label{fig1b}
	\end{subfigure}
	\caption{Configuration of problem for two different polarizations of the laser beam (a)parallel and (b)perpendicular polarization.}
	\label{fig1}
\end{figure}
One can show that the total internal electric field related to the conduction electrons of MNP vanishes, thus we write the following momentum equation for the displacement of center of mass of conduction electrons of the first NP, ${{\bf{r}}_1}$, which in the rigid body approximation is equivalent to the displacement of electronic cloud of each NP from the equilibrium state \cite{kheirandish2019modified}
\begin{align}
	{m_e}\frac{{{d^2}{{\bf{r}}_1}}}{{d{t^2}}} + {m_e}\gamma \frac{{d{{\bf{r}}_1}}}{{dt}} + {m_e}\xi \omega _p^2{{\bf{r}}_1} =  - e{{\bf{E}}_1},
\end{align}
where ${m_e}$, $e$, $\gamma $ and $\xi $ are the electron mass, the magnitude of electron charge, damping coefficient and a function of NP radius obtained from the experimental data \cite{kheirandish2019modified}, respectively, ${\omega _p} = \sqrt {{n_0}{e^2}/({m_e}{\varepsilon _0})} $ is plasma frequency of conduction electrons, ${n_0}$ is the density of conduction electrons and ${\varepsilon _0}$ is the vacuum permittivity. It is better to mention that the third term of the left side in Eq. (2) is related to the restoration force caused by the separation of electrons from the positive background ions and parameter $\xi $ is for adjustment of experimental data with the theory. In most of theoretical studies related to the interaction of laser beam with MNPs \cite{sepehri2015self,sepehri2015raman,sepehri2017magnetic,sepehri2015nonlinear,sepehri2016dielectric,javan2019magnetic,javan2019semi,kheirandish2018polarization}, in an ideal case of rigid body approximation in which all conduction electrons are replaced equally, its value is $1/3$, however in the real situation, its value depending on the diameter of NP is less than $1/3$ and it vanishes for the large particles which shows the transition to bulk medium situation \cite{kheirandish2019modified}. ${{\bf{E}}_1}$ is the electric field at the place of the first NP which is the sum of the incident wave electric field ${{\bf{E}}_L}$ and interaction term ${{\bf{E}}_{2 \to 1}}$ where ${{\bf{E}}_{2 \to 1}}$ is the electric field at the place of first NP ${{\bf{r}}_{01}}$ caused by the electric dipole moment of the second NP (${{\bf{p}}_2} =  - Ze{{\bf{r}}_2}$, where $Z = {n_0}V$ is the total conduction electrons of each NP and $V$ is the volume of NP) that can be represented as
\begin{align}
	&{{\bf{E}}_{2 \to 1}} = \frac{1}{{4\pi {\varepsilon _0}}}\left[ {\frac{{3{{{\bf{\hat e}}}_2}({{\bf{p}}_2}.{{{\bf{\hat e}}}_2}) - {{\bf{p}}_2}}}{{{d^3}}}} \right. \nonumber \\
	&\left. {\,\,\,\,\,\,\,\, - ik\frac{{3{{{\bf{\hat e}}}_2}({{\bf{p}}_2}.{{{\bf{\hat e}}}_2}) - {{\bf{p}}_2}}}{{{d^2}}}\, + {k^2}\frac{{{{\bf{p}}_2} - {{{\bf{\hat e}}}_2}({{\bf{p}}_2}.{{{\bf{\hat e}}}_2})}}{d}} \right],
\end{align}
where ${{\bf{\hat e}}_2}$ is an unit vector whose orientation is from the center of the first NP to the center of other one. Substituting Eqs. (1) and (3) into Eq. (2), we obtain the following equation for the displacement of electrons of first NP
\begin{align}
	\frac{{{d^2}{z_1}}}{{d{t^2}}} + \gamma \frac{{d{z_1}}}{{dt}} + \xi \omega _p^2{z_1} =&  - \frac{{eE}}{{2{m_e}}}{e^{i(kx - \omega t)}}\nonumber \\
	 &+ \frac{{Z{e^2}}}{{2\pi {\varepsilon _0}{m_e}}}(\frac{1}{{{d^3}}} - \frac{{ik}}{{{d^2}}}){z_2},
\end{align}
By interchanging between indeces 1 and 2, one can obtain other equation for the displacement of electronic cloud of the second NP as following
\begin{align}
	\frac{{{d^2}{z_2}}}{{d{t^2}}} + \gamma \frac{{d{z_2}}}{{dt}} + \xi \omega _p^2{z_2} =&  - \frac{{eE}}{{2{m_e}}}{e^{i(kx - \omega t)}}\nonumber \\
	 &+ \frac{{Z{e^2}}}{{2\pi {\varepsilon _0}{m_e}}}(\frac{1}{{{d^3}}} - \frac{{ik}}{{{d^2}}}){z_1},
\end{align}
Considering periodic solutions of ${z_1} = (1/2){\tilde z_1}\,{{\mathop{\rm e}\nolimits} ^{i(kx - \omega t)}}$ and ${z_2} = (1/2){\tilde z_2}\,{{\mathop{\rm e}\nolimits} ^{i(kx - \omega t)}}$ for Eqs. (4) and (5), the amplitude of displacement of each NP electronic cloud is achieved as
\begin{align}
	{\tilde z_1} = {\tilde z_2} = \frac{1}{{1 - \alpha }}{z_0},
\end{align}
where
\begin{align}
	{z_0} = \frac{{eE}}{{{m_e}({\omega ^2} + i\omega \gamma  - \xi \omega _p^2)}},
\end{align}
and
\begin{align}
	\alpha  = \frac{{ - Z{e^2}}}{{2\pi {\varepsilon _0}{m_e}{d^3}({\omega ^2} + i\omega \gamma  - \xi \omega _p^2)}}\left( {1 - ikd} \right),
\end{align}
By defining the polarization vector as ${\bf{P}} = {\varepsilon _0}\chi {{\bf{E}}_L}$, we can obtain the electric susceptibility as
\begin{align}
	\chi  = \left( {\frac{1}{{1 - \alpha }}} \right)\frac{{ - \omega _p^2}}{{{\omega ^2} + i\omega \gamma  - \xi \omega _p^2}}.
\end{align}
Eventually, we can derive the electric permittivity for two interactional nanoparticles as:
\begin{align}
	{\left( {\frac{\varepsilon }{{{\varepsilon _0}}}} \right)_{\,{\rm{||}}}} &= n_{NP}^2 = 1 + \chi \nonumber \\
	&= 1 - \left( {\frac{1}{{1 - \alpha }}} \right)\frac{{\omega _p^2}}{{{\omega ^2} + i\omega \gamma  - \xi \omega _p^2}},
\end{align}
where ${n_{NP}}$ is the refractive index of each NP. It is worth mentioning that in Drude-Lorentz model, in order to consider the role of bound electrons in the permittivity, one should change Eq. (10) as following
\begin{align}
{\left( {\frac{\varepsilon }{{{\varepsilon _0}}}} \right)_{\,{\rm{||}}}} = {\varepsilon _\infty } - \left( {\frac{1}{{1 - \alpha }}} \right)\frac{{\omega _p^2}}{{{\omega ^2} + i\omega \gamma  - \xi \omega _p^2}},
\end{align}
where ${\varepsilon _\infty }$ is determined through experimental data of bulk medium \cite{johnson1972optical}.\\
For the small values of $\alpha $, expanding the term ${(1 - \alpha )^{ - 1}}$ with respect to $\alpha $ and saving only its linear terms in Eq. (11), we get
\begin{align}
	{\left( {\frac{\varepsilon }{{{\varepsilon _0}}}} \right)_{\,{\rm{||}}}} = {\left( {\frac{\varepsilon }{{{\varepsilon _0}}}} \right)_{{\rm{single}}}} - {\left( {\frac{{\Delta \varepsilon }}{{{\varepsilon _0}}}} \right)_{{\rm{int,}}\,{\rm{||}}}},
\end{align}
where the first term is the relative permittivity of a single non-interactional NP \cite{kheirandish2019modified}
\begin{align}
	{\left( {\frac{\varepsilon }{{{\varepsilon _0}}}} \right)_{{\rm{single}}}} = {\varepsilon _\infty } - \frac{{\omega _p^2}}{{{\omega ^2} + i\omega \gamma  - \xi \omega _p^2}},
\end{align}
and the second term is its deviation due to the interaction
\begin{align}
	{\left( {\frac{{\Delta \varepsilon }}{{{\varepsilon _0}}}} \right)_{{\rm{int,}}\,\,{\rm{||}}}} = \frac{{\alpha \omega _p^2}}{{{\omega ^2} + i\omega \gamma  - \xi \omega _p^2}}.
\end{align}
\\
%********************************************
\subsection{Perpendicular Polarization}
%********************************************

Now, let us consider another different laser beam electric field polarization where it is perpendicular to the NPs symmetry axis or equivalently its magnetic field is parallel with the orientation of dimer axis as it is shown in figure (1-b). By a similar method which is engaged in the previous sub-section, we solve the motion equations. For brevity, we will discard repetitive details and only bring the final results. For this case, according to figure (1-b), we take the laser fields as following
\begin{align}
	{{\bf{E}}_L} = \frac{1}{2}E{e^{i(kx - \omega t)}}{{\bf{\hat e}}_y} + c.c.,
\end{align}
which leads to the following equations for the amplitudes of the displacements of each NP
\begin{align}
	{\tilde y_1} = {\tilde y_2} = \frac{1}{{1 - \beta }}{y_0},
\end{align}
where ${y_0} = {z_0}$ and
\begin{align}
	\beta  = \frac{{Z{e^2}(1 - ikd - {k^2}{d^2})}}{{4\pi {\varepsilon _0}{m_e}{d^3}({\omega ^2} + i\omega \gamma  - \xi \omega _p^2)}},
\end{align}
After some simple algebraic operations, we get the following equation for the relative permittivity of each particle
\begin{align}
	{\left( {\frac{\varepsilon }{{{\varepsilon _0}}}} \right)_ \bot } = n_{NP}^2 = 1 - \left( {\frac{1}{{1 - \beta }}} \right)\frac{{\omega _p^2}}{{{\omega ^2} + i\omega \gamma  - \xi \omega _p^2}},
\end{align}
Finally, for the small values of $\beta $,  from Eq. (18), we obtain the relative permittivity for two interactional NPs as
\begin{align}
	{\left( {\frac{\varepsilon }{{{\varepsilon _0}}}} \right)_ \bot } = {\left( {\frac{\varepsilon }{{{\varepsilon _0}}}} \right)_{{\rm{single}}}} - {\left( {\frac{{\Delta \varepsilon }}{{{\varepsilon _0}}}} \right)_{{\rm{int,}} \bot }},
\end{align}
where
\begin{align}
	{\left( {\frac{{\Delta \varepsilon }}{{{\varepsilon _0}}}} \right)_{{\rm{int,}} \bot }} = \frac{{\beta \omega _p^2}}{{{\omega ^2} + i\omega \gamma  - \xi \omega _p^2}}.
\end{align}
Comparison between Eqs. (12) and (19) shows that there is substantial difference between two different laser beam polarizations, especially when $kd$ is comparable with the unity.\\

\begin{figure}
	\begin{subfigure}{0.40\textwidth}\includegraphics[width=\textwidth]{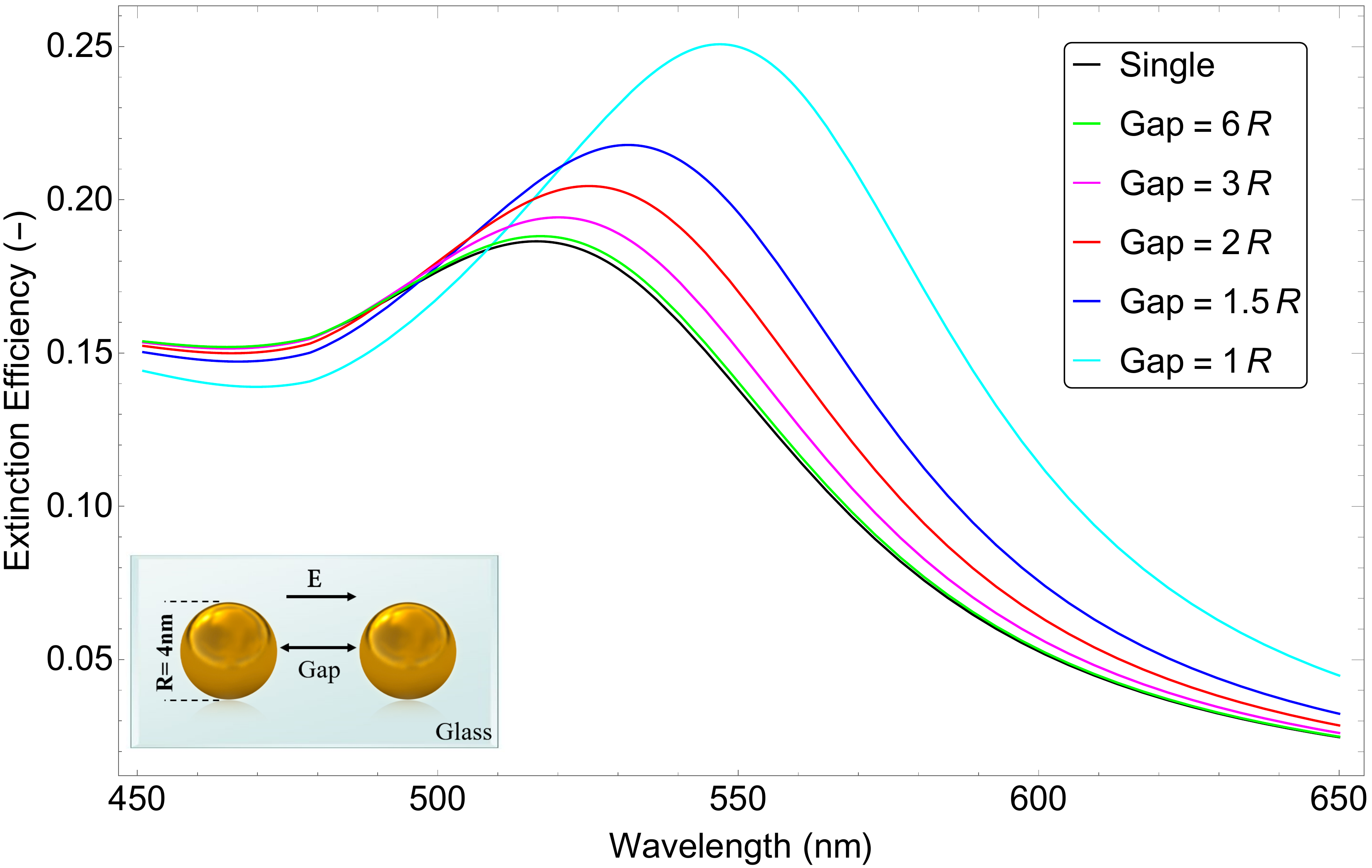}
		\caption{}
		\label{fig2a}
	\end{subfigure}
	\begin{subfigure}{0.40\textwidth}\includegraphics[width=\textwidth]{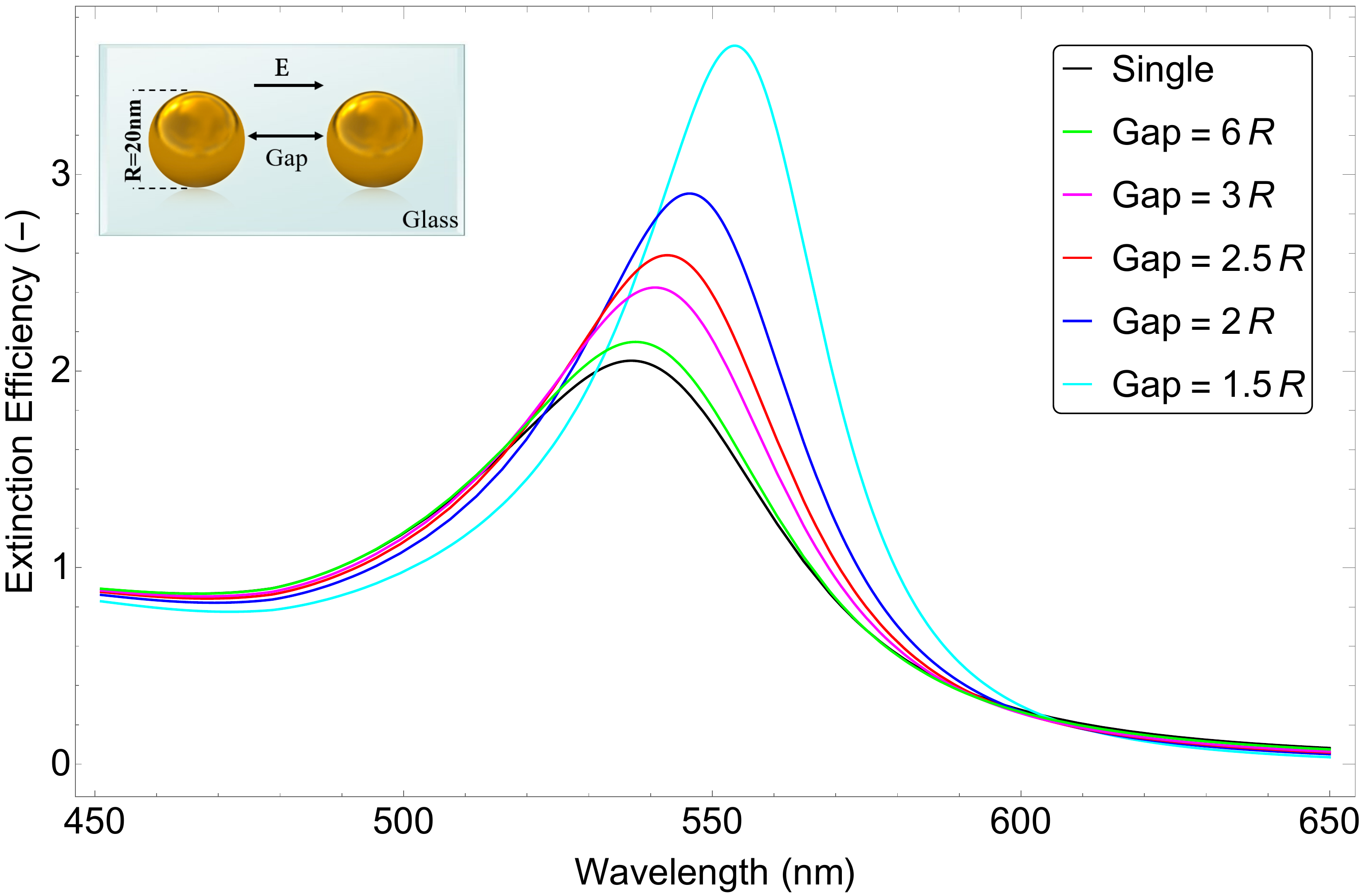}
		\caption{}
		\label{fig2b}
	\end{subfigure}
	\begin{subfigure}{0.40\textwidth}\includegraphics[width=\textwidth]{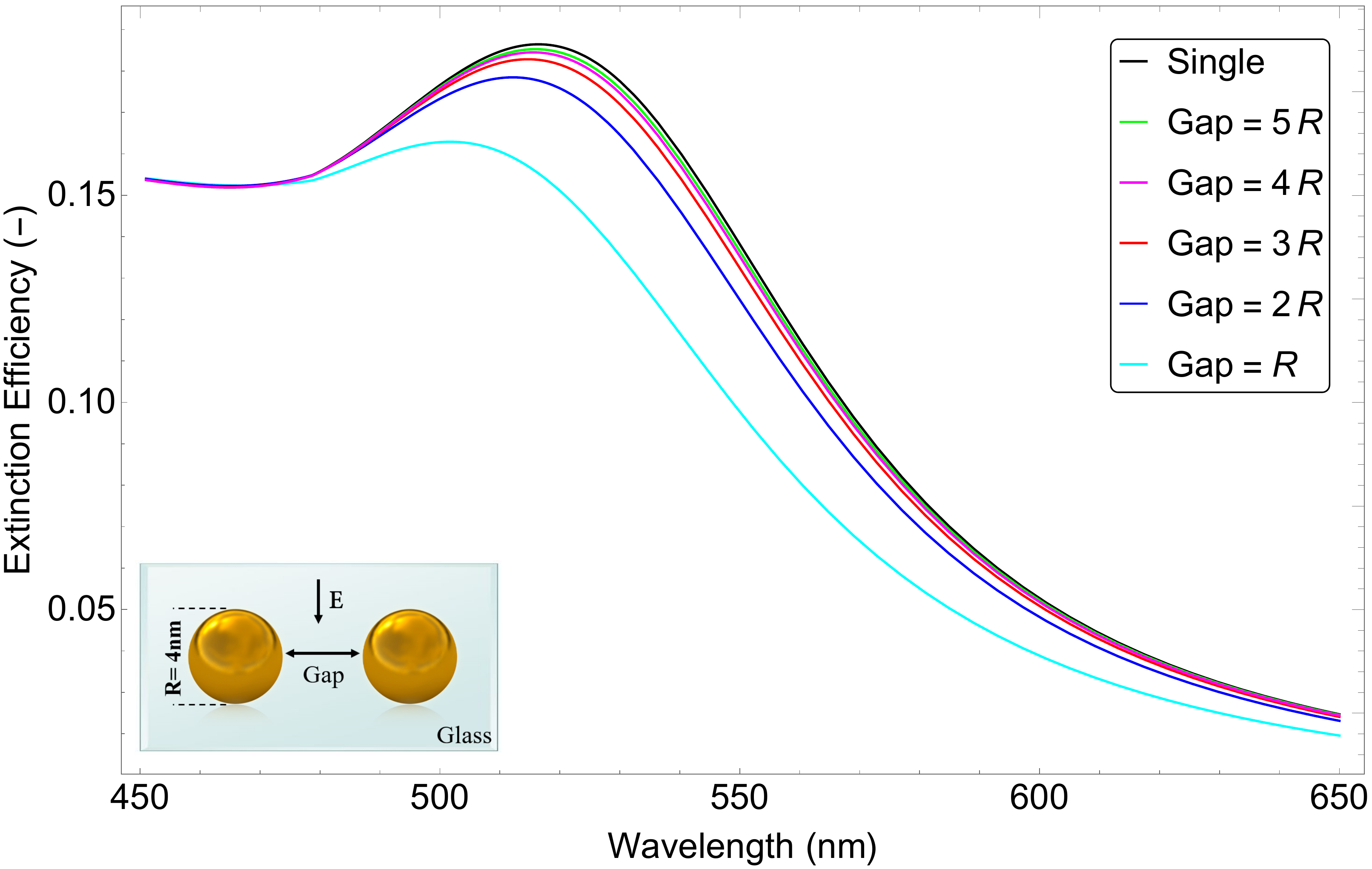}
		\caption{}
		\label{fig2c}
	\end{subfigure}
	\begin{subfigure}{0.40\textwidth}\includegraphics[width=\textwidth]{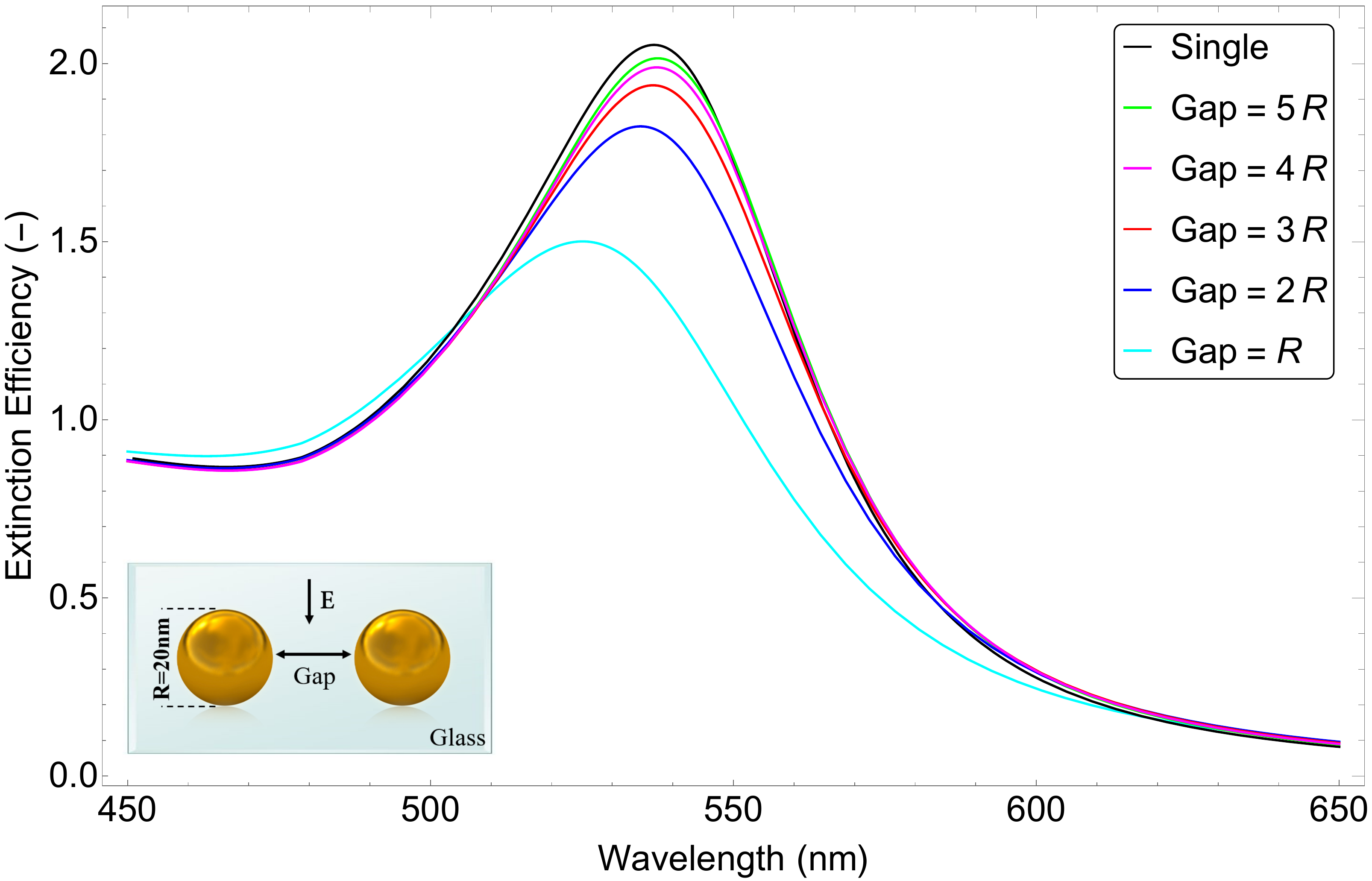}
		\caption{}
		\label{fig2d}
	\end{subfigure}
	\caption{Extinction efficiency spectra of gold NP dimer with diameters of (a), (c) $4nm$ and (b), (d) $20nm$, with different interparticle separation gaps for two different polarizations of (a), (b) parallel and (c), (d) perpendicular.}
	\label{fig2}
\end{figure}
%********************************************
\section{NUMERICAL ANALYSIS}
%********************************************

In order to show the interaction effect between the nanoparticles, we perform some numerical investigations. In experiments, the measurable important parameter reflecting the linear optical properties of NP is the extinction cross section which is defined as the sum of scattering and absorption cross sections according to the following equation extracted from the well-known Mie theory \cite{bohren2008absorption}
\begin{align}
	{C_{ext}} = \frac{{2\pi }}{{{k^2}}}\sum\limits_{n = 1}^\infty  {(2n + 1){\mathop{\rm Re}\nolimits} ({a_n} + {b_n})},
\end{align}
where
\begin{align}
	{a_n} = \frac{{m{\psi _n}(mx){{\psi '}_n}(x) - {\psi _n}(x){{\psi '}_n}(mx)}}{{m{\psi _n}(mx){{\xi '}_n}(x) - {\xi _n}(x){{\psi '}_n}(mx)}}, \nonumber \\
	{b_n} = \frac{{{\psi _n}(mx){{\psi '}_n}(x) - m{\psi _n}(x){{\psi '}_n}(mx)}}{{{\psi _n}(mx){{\xi '}_n}(x) - m{\xi _n}(x){{\psi '}_n}(mx)}},
\end{align}
and $x = 2\pi {n_b}a/\lambda $ is the size parameter, $m = {n_{NP}}/{n_b}$ is the relative refractive index, where ${n_b}$ is the refractive index of the background medium, ${\psi _n}(x)$ and ${\xi _n}(x)$ are Riccati-Bessel functions \cite{bohren2008absorption}.\\
In Figs. (2-a)-(2-d), the extinction efficiency (i.e., extinction cross sections normalized by the particle cross section ${C_{ext}}/\pi {a^2}$) of two interactional gold NPs are presented with respect to the variations of laser beam wavelength for different interparticle separation gaps, different sizes of NPs and two different laser beam polarizations as well. In Figs. (2-a) and (2-c), we have considered a pair of NPs with diameter of $R = 4nm$ that are doped in a glass background medium while Figs. (2-b) and (2-d) correspond to a NP dimer with diameter of $R = 20nm$. For $R = 4nm$, the data of single particle permittivity parameters, i.e. $\gamma $ and $\xi $, have been taken from our previous work \cite{kheirandish2019modified} and for $R = 20nm$ these parameter are determined in the appendix. For comparison, we have plotted the extinction efficiency of a single NP when it does not interact with other NPs. In Figs. (2-a) and (2-b), in a fixed size of NP where the laser beam electric field polarization is parallel to the NPs symetry axis, decrease in the gap causes increase in the extinction cross section value at SPR wavelength region. Also the place of SPR wavelength experiences a red-shift via interaction of particles. Such results are taken in Refs. \cite{jain2007universal,jain2010plasmonic,rechberger2003optical} where the extinction cross section of identical gold discs is achieved experimentally. In contrast with the parallel polarization, in Figs. (2-c) and (2-d), where the laser beam electric field is perpendicular to the dimer axis, we can see that the interparticle interaction leads to the decrease in the extinction efficiency at the SPR region. In this case, interaction of two NPs results in a blue-shift. These results are also in agreement with the experimental results of Refs \cite{jain2007universal,jain2010plasmonic,rechberger2003optical} for gold nanodiscs.\\
\begin{figure}
	\begin{subfigure}{0.40\textwidth}\includegraphics[width=\textwidth]{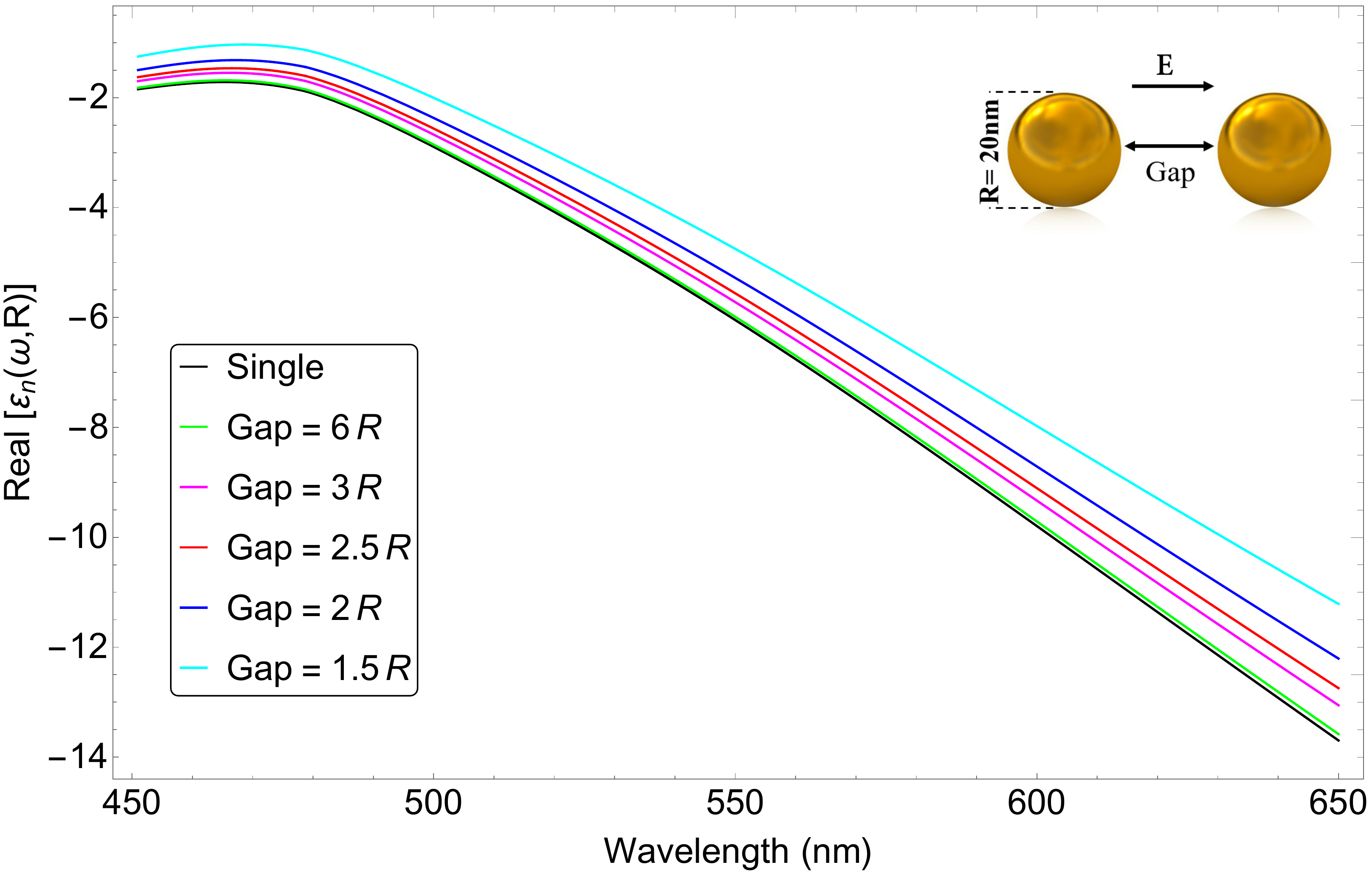}
		\caption{}
		\label{fig3a}
	\end{subfigure}
	\begin{subfigure}{0.40\textwidth}\includegraphics[width=\textwidth]{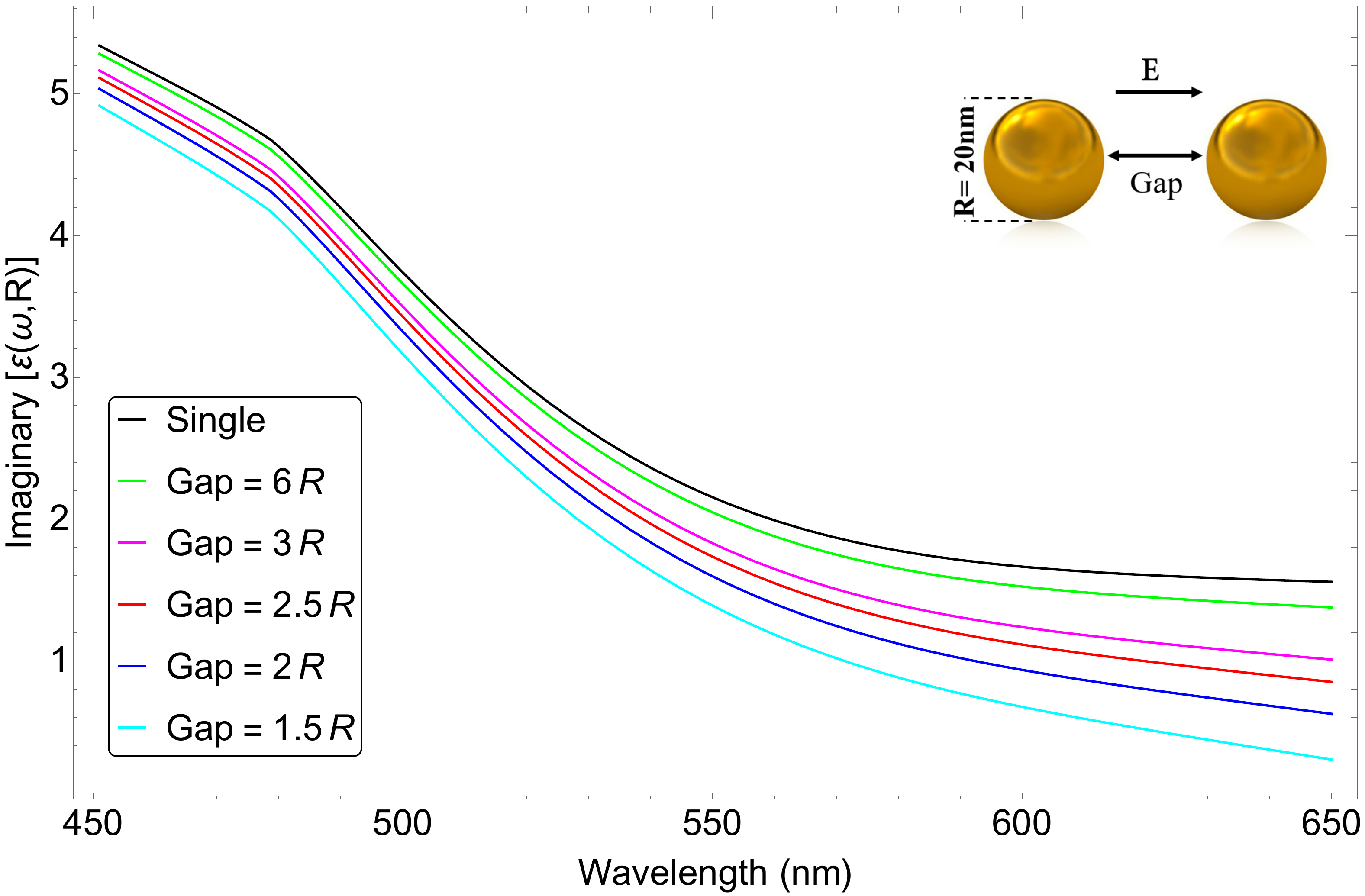}
		\caption{}
		\label{fig3b}
	\end{subfigure}
	\begin{subfigure}{0.40\textwidth}\includegraphics[width=\textwidth]{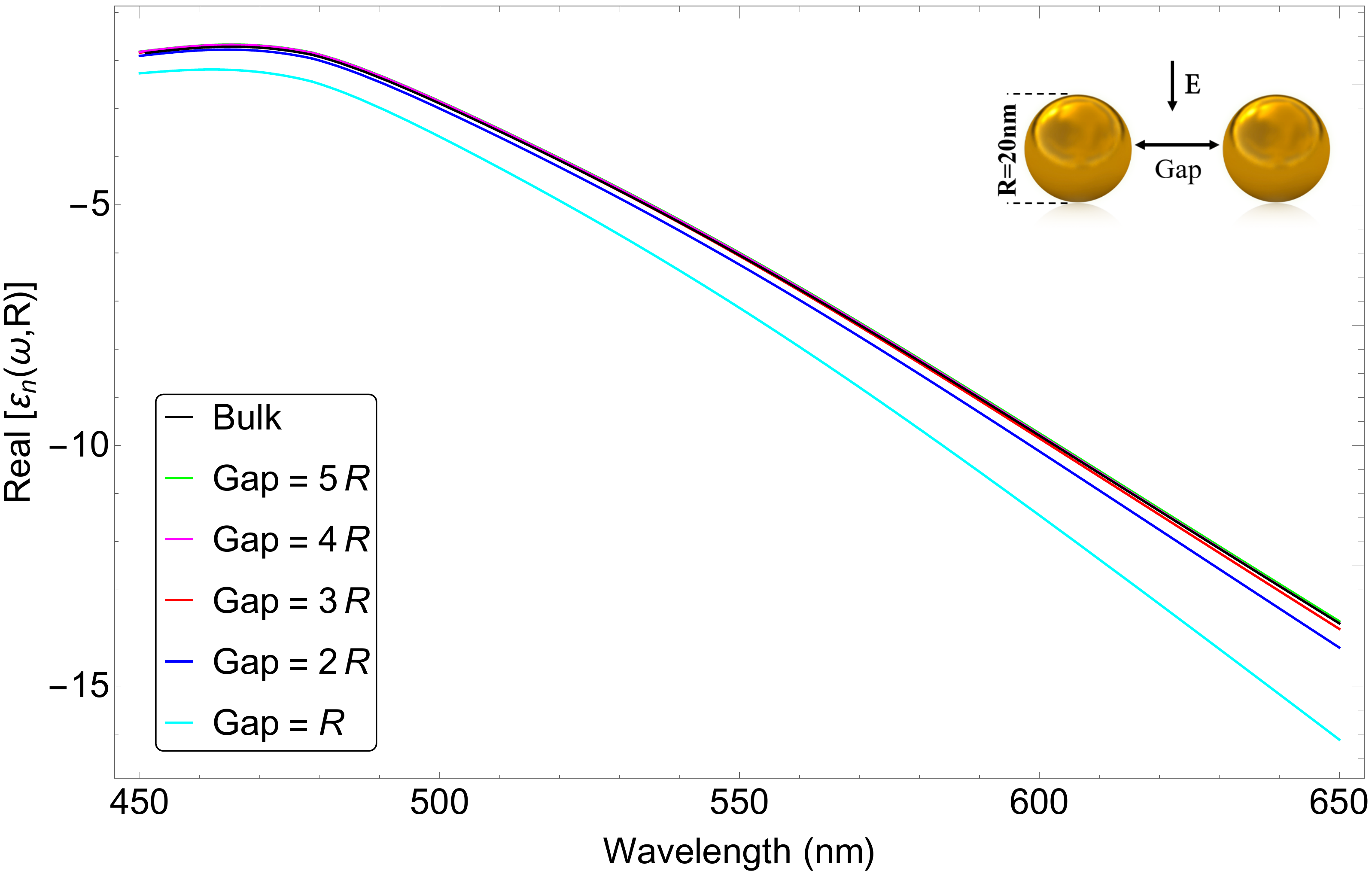}
		\caption{}
		\label{fig3c}
	\end{subfigure}
	\begin{subfigure}{0.40\textwidth}\includegraphics[width=\textwidth]{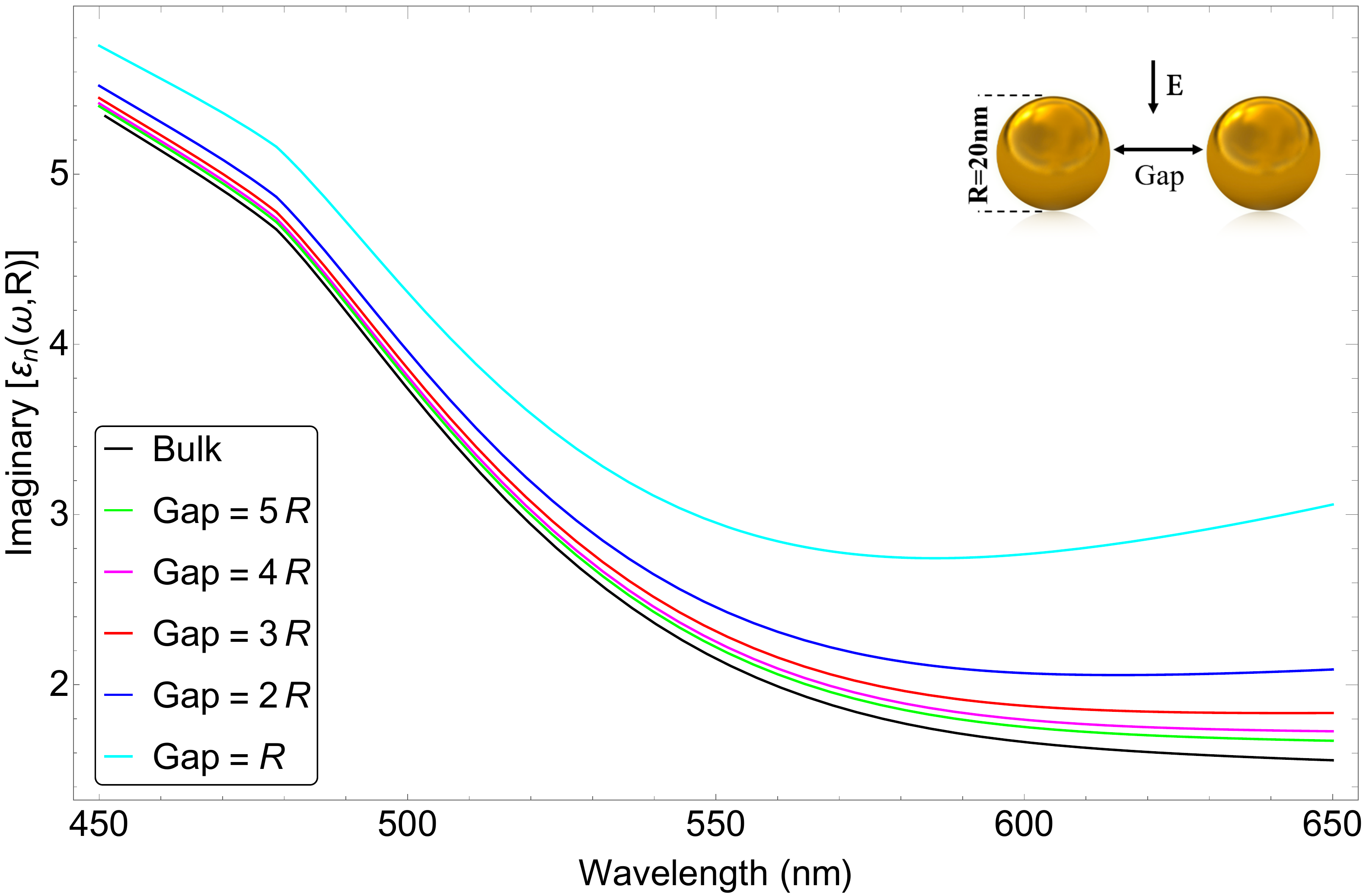}
		\caption{}
		\label{fig3d}
	\end{subfigure}
	\caption{Variations of (a), (c) real and (b), (d) imaginary parts of permittivity as a function of wavelength for gold NP dimer with diameters of $20nm$, with different interparticle separation gaps for two different polarizations of (a), (b) parallel and (c), (d) perpendicular.}
	\label{fig3}
\end{figure}
\begin{figure}
	\begin{subfigure}{0.40\textwidth}\includegraphics[width=\textwidth]{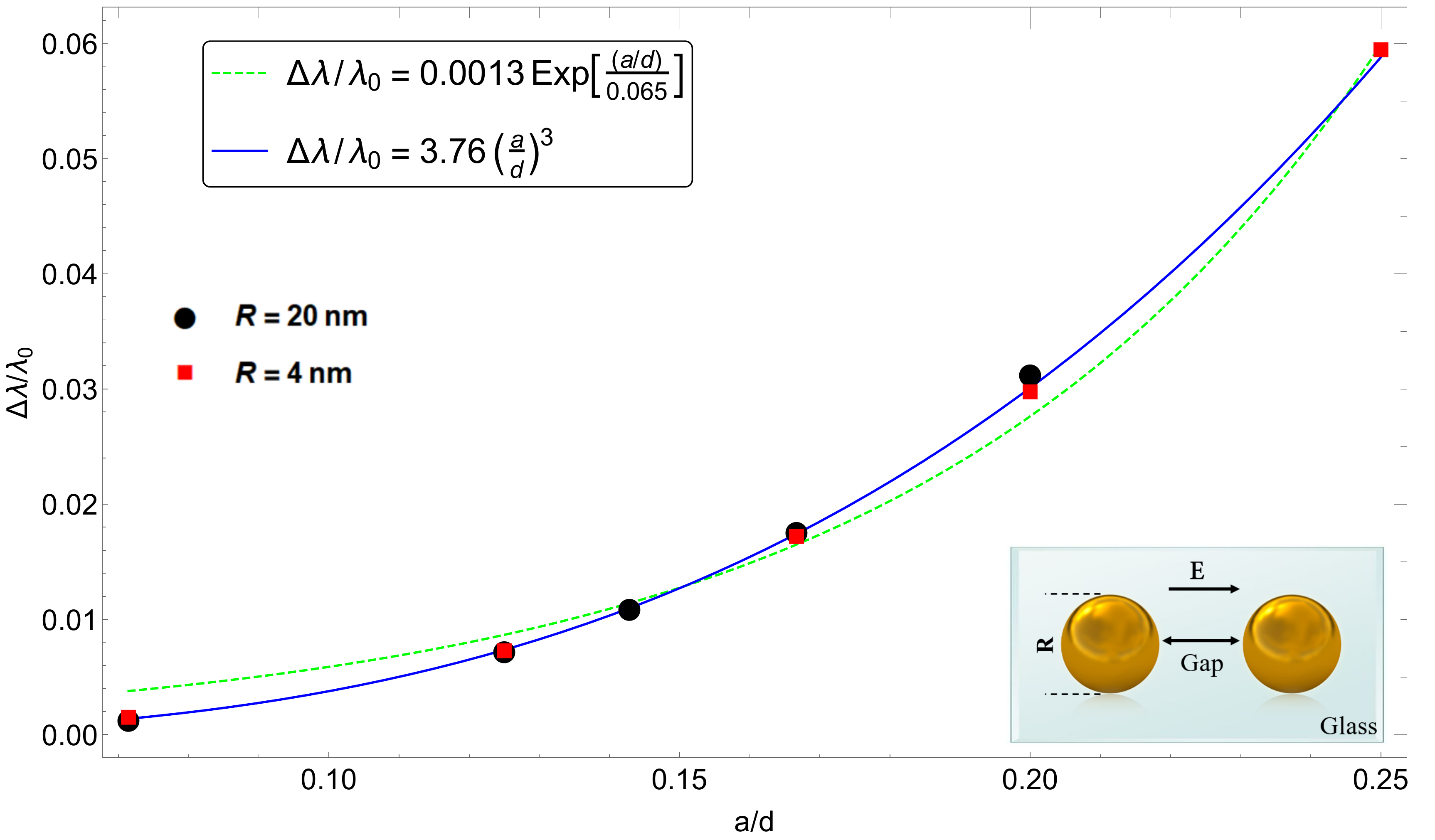}
		\caption{}
		\label{fig4a}
	\end{subfigure}
	\begin{subfigure}{0.40\textwidth}\includegraphics[width=\textwidth]{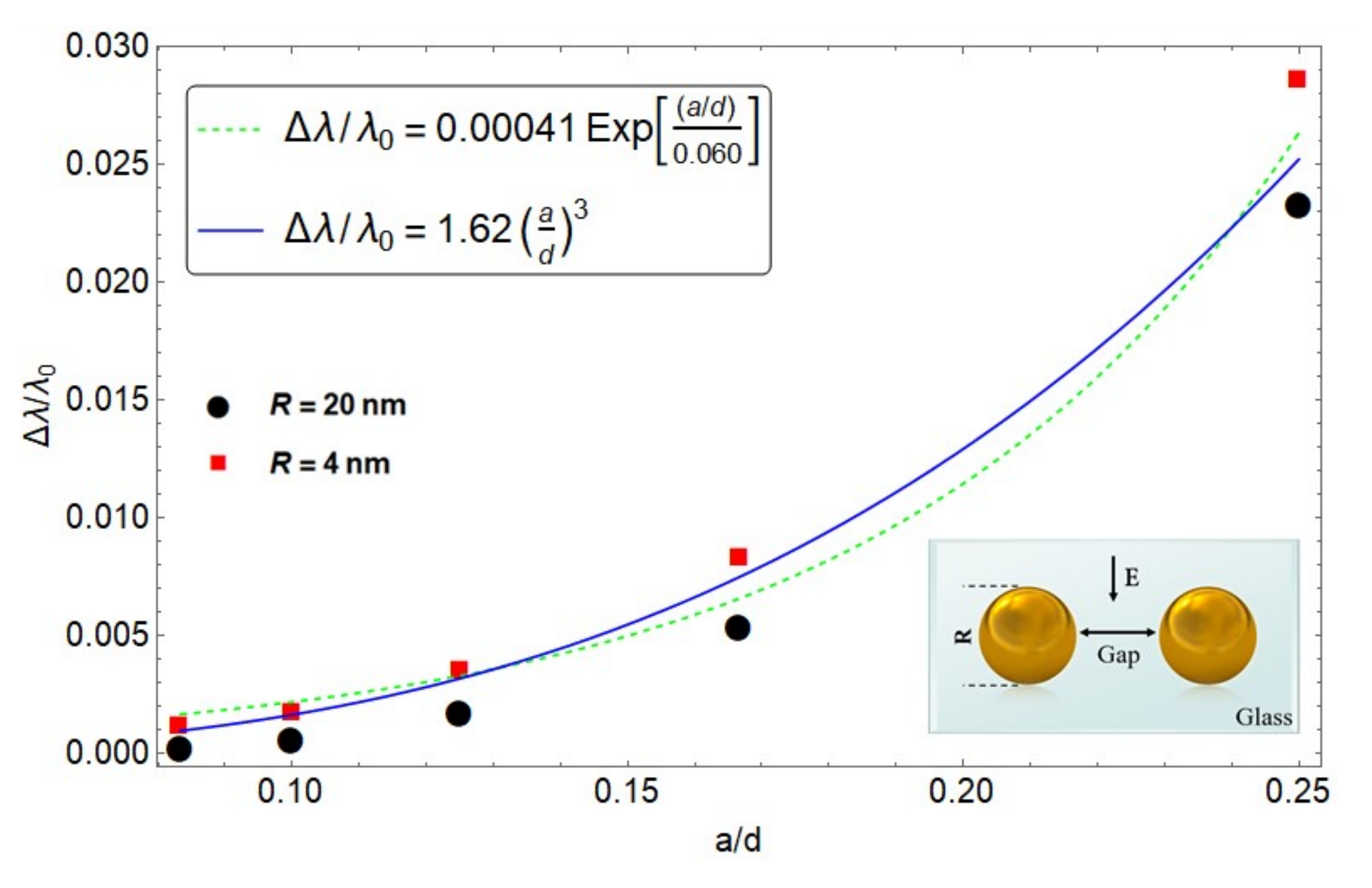}
		\caption{}
		\label{fig4b}
	\end{subfigure}
	\caption{Shift in the SPR wavelength of a pair of gold NPs as a function of the geometric parameter ($a/d$).}
	\label{fig1}
\end{figure}
In Figs. (3-a)-(3-d), the variations of real and imaginary parts of permittivity have been plotted as a function of wavelength for different interparticle separation gaps for the NPs diameter of $R = 20nm$. In Figs. (3-a) and (3-b) the electric field is taken parallel with the orientation of dimer axis while it is taken perpendicular in the Figs. (3-c) and (3-d). As one can see, in Figs. (3-a) and (3-b) interparticle interactions cause the decrease in the absolute value of real and imaginary parts. In opposite, for the perpendicular polarization, we can see from Figs. (3-c) and (3-d) that decrease in the interparticle separation gap of NPs causes the increase in the absolute value of the real and imaginary parts of the permittivity. This different behavior in the effect of interaction on the permittivity can be considered as the source of different behavior of extinction cross section with respect to the different polarizations.\\
Figures (4-a) and (4-b) show the variations of the relative shift of SPR wavelength $\Delta \lambda /{\lambda _0}$ versus the ratio of NP radius to the separation $a/d$ for different polarizations and different NP radii where $\Delta \lambda  = \left| {{\lambda _{\max }} - {\lambda _0}} \right|$, ${\lambda _0}$ is the single NP SPR wavelength and ${\lambda _{\max }}$ is the gold NP dimer SPR wavelength. In previous experimental studies for gold nanodisc dimers \cite{rechberger2003optical}, it is shown that the relationship between $\Delta \lambda /{\lambda _0}$ and $a/d$ can be fitted by an exponential function. Here also for all cases, we have shown that this exponential relation can be established, however there is another cubic functionality of SPR wavelength displacement with respect to the radius to separation ratio. For NP diameters of $R = 4nm$, $R = 20nm$ and parallel polarization, in Fig (4-a), the SPR wavelength fits by through the following functions
\begin{align}
	{\left( {\frac{{\Delta \lambda }}{{{\lambda _0}}}} \right)_{{\rm{||}}}} = 0.0013\exp \,\left( {\frac{{a/d}}{{0.065}}} \right),
\end{align}
\begin{align}
	{\left( {\frac{{\Delta \lambda }}{{{\lambda _0}}}} \right)_{{\rm{||}}}} = 3.76\,\,{\left( {\frac{a}{d}} \right)^3},
\end{align}
As fitting results show, here, the cubic functionality is better than the exponential one. Origin of the cubic functionality is located in the dipole-dipole interaction terms. The dipole-dipole terms for the parallel polarization case include ${(a/d)^3}$ and ${a^3}/(\lambda {d^2})$ where for the rigid body approximation criterion $a/\lambda  <  < 1$, the former is dominant. In the case of perpendicular polarization, besides two mentioned terms in the dipole-dipole interaction, the term ${a^3}/(d{\lambda ^2})$ appears as well and in this case for $a/\lambda  <  < 1$ the more effective interactional term again is ${(a/d)^3}$. In Fig. (4-b), for the same sizes of dimers but for the perpendicular polarization, the two best fitted functions are
\begin{align}
	{\left( {\frac{{\Delta \lambda }}{{{\lambda _0}}}} \right)_ \bot } = 0.00041\exp \,\left( {\frac{{a/d}}{{0.060}}} \right),
\end{align}
\begin{align}
	{\left( {\frac{{\Delta \lambda }}{{{\lambda _0}}}} \right)_ \bot } = 1.62\,\,{\left( {\frac{a}{d}} \right)^3},
\end{align}
here also the cubic function better fits and the dipole-dipole interaction nature of the plasmon resonance shift is more evident. \\

%********************************************
\section{CONCLUSIONS}
%********************************************

In this theoretical investigation, the effect of dipole-dipole interaction on the linear optical properties of a MNP dimer is studied. Analytical expressions are derived for the permittivity of interactional particles through a Drude-like model where the classical confinement effect is considered by taking into account the restoration force whose proper value is determined through experimental data. Calculations are accomplished for two different cases of laser beam polarizations where the orientation of electric field of laser beam is parallel or perpendicular to the symmetry axis of the dimer. It is found that for the parallel case, interaction of MNPs causes the increase in the height of maximum value of extinction cross section related to the SPR. Also, it leads to the red-shift of SPR wavelength with respect to the non-interactional individual NP. For the perpendicular polarization, we get the opposite effect where the interaction results in the decrease in the plasmon resonance peak and the blue-shift of its wavelength. Our analytical results are in agreement with the experimental ones. Finally, the relationship between SPR wavelength and geometric factor $a/d$ has been investigated for different situations and it is shown that it can be fitted not only by an exponential function nor by a cubic one which reflects the dominant dipole-dipole characteristic of the interaction. \\

\begin{figure}[b]
	\centerline{\includegraphics[scale=.25]{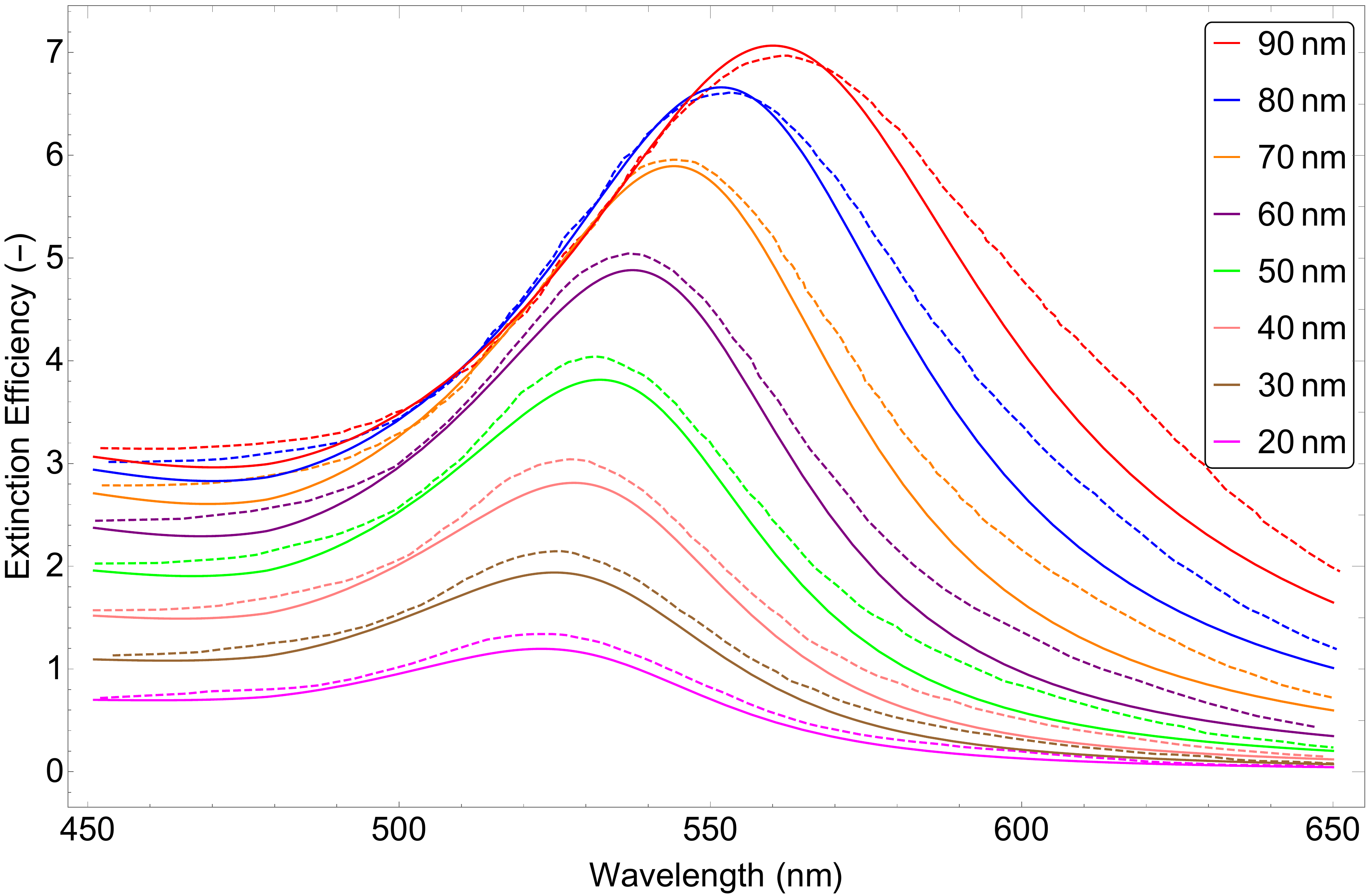}}
	\caption{The calculated extinction efficiency (solid lines) and the experimentally measured ones (dotted line) in dependence of wavelength for NPs diameters of $20$, $30$, $40$, $50$, $60$, $70$, $80$, $90nm$ (the order of diameter increase is from bottom to top).}
	\label{fig5}
\end{figure}
%********************************************
\section{APPENDIX}
%********************************************

%********************************************
\subsection{Single Nanoparticles}
%********************************************

In a recent study \cite{kheirandish2019modified}, we have shown that based on a phenomenological Drude-like model which includes the role of classical confinement related to the appearance of restoration force caused by the displacement of conduction electrons with respect to the positive ionic background,  the permittivity of a single NP can be determined properly through extracting free phenomenological parameters of $\xi $ and $\gamma $ by experimental data of extinction cross section of individual NPs. For small NPs these parameters are considered as a function of wavelength and NP radius. In this appendix, we use the same approach in order to find permittivity of large gold NPs ranged from $20nm$ to $90nm$ whose extinction cross section can be found in experimental investigations \cite{starowicz2018tuning,hong2013optimal}. The value of the free parameters $\xi $ and $\gamma $ in the permittivity of an NP with a given diameter are suggested through a trial and error process in order to by applying this permittivity in the extinction cross section of Eq. (20), good agreement reveals between experimental and theoretical data.  In Fig. (5), the extinction efficiency of a single gold NP suspended in a water medium has been plotted for different sizes of spherical NPs. The dotted curves are obtained experimentally \cite{starowicz2018tuning} and the solid lines show our model results. For the model, we have used the phenomenological parameters as follows
\begin{align}
	\xi  = {\xi _0} + {c_1}\frac{1}{a} + {c_2}\frac{1}{{{a^2}}} + {c_3}\frac{1}{{{a^3}}},
\end{align}
where
\begin{align}
	&{\xi _0} =  - 0.950 \times {10^{ - 3}},\,\,\,{c_1} = 5.44nm,\,\nonumber \\
	&{c_2} =  - 1.17n{m^2},\,\,\,{c_3} = 0.08n{m^3},
\end{align}
and $a$ is in nm. For the damping factor $\gamma $ of NP, we use the following expression
\begin{align}
	\gamma  = {\gamma _0} + {\gamma _{surf}},
\end{align}
where ${\gamma _0} = 0.7 \times {10^{14}}{s^{ - 1}}$ and ${\gamma _{surf}} = A{v_F}/a$ is the electron damping coefficient caused by the scattering of electron by the surface of NP which is called the mean free path limitation effect \cite{genzel1975dielectric,coronado2003surface}, here $A = 0.25$ is a dimensionless parameter and ${v_F}$ is the gold Fermi velocity \cite{berciaud2005observation}.

\section{Reference}
\bibliography{refs}
\end{document}